\documentclass[12pt]{article}

\usepackage{amsfonts}
\usepackage{amsmath}
\usepackage{amssymb}
\usepackage[dvips]{graphicx}
\usepackage{comment}

\newcommand{\ba}{\begin{eqnarray}}
\newcommand{\ea}{\end{eqnarray}}
\newcommand{\bi}{\begin{itemize}}
\newcommand{\ei}{\end{itemize}}
\newcommand{\nn}{\nonumber}

\newcommand{\ut}{\underline{t}}
\newcommand{\us}{\underline{s}}
\newcommand{\ua}{\underline{\alpha}}
\newcommand{\ur}{\underline{\rho}}
\newcommand{\ttg}{\mathtt{g}}
\newcommand{\tth}{\mathtt{h}}
\newcommand{\ttk}{\mathtt{k}}
\newcommand{\cF}{\mathcal{F}}

\newcommand{\tA}{\tilde{A}}
\newcommand{\ta}{\tilde{a}}
\newcommand{\tB}{\tilde{B}}
\newcommand{\tF}{\tilde{F}}
\newcommand{\tZ}{\tilde{\mathcal{Z}}}
\newcommand{\tL}{\tilde{\Lambda}}
\newcommand{\tO}{\tilde{\Omega}}
\newcommand{\bA}{\mathbf{A}}
\newcommand{\bF}{\mathbf{F}}
\newcommand{\bL}{\mathbf{\Lambda}}
\newcommand{\bD}{\mathbf{D}}

\newcommand{\Tr}{\mathrm{Tr}}

\begin{document}

\begin{titlepage}

\begin{center}

\begin{flushright}
UT-12-14
\end{flushright}

\vskip 12mm

\textbf{\LARGE Note on non-Abelian two-form gauge fields}

\vskip 2cm
{\large
Pei-Ming Ho$^\dagger$\footnote{
e-mail address: pmho@phys.ntu.edu.tw} and 
Yutaka Matsuo$^\ddagger$\footnote{
e-mail address:
matsuo@phys.s.u-tokyo.ac.jp}}\\
\vskip 2cm
{\it\large
$^\dagger$
Department of Physics and Center for Theoretical Sciences, \\
Center for Advanced Study in Theoretical Sciences, \\
National Center for Theoretical Sciences, \\
National Taiwan University, Taipei 10617, Taiwan,
R.O.C.}\\
\vskip 3mm
{\it\large
$^\ddagger$
Department of Physics, Faculty of Science, University of Tokyo,\\
Hongo 7-3-1, Bunkyo-ku, Tokyo 113-0033, Japan\\
\noindent{ \smallskip }\\
}
\vspace{60pt}
\end{center}
\begin{abstract}

Motivated by application to multiple M5-branes, we study
some properties of non-Abelian two-form gauge theories.
We emphasize that the fake curvature condition which is commonly
used in the literature
would restrict the dynamics to be either a free theory or a topological system.
We then propose a modification of transformation law which
simplifies the gauge transformation of 3-form field strength and
enables us to write down a gauge invariant action.
We then argue that a generalization of Stueckelberg mechanism
naturally gives mass to the two-form gauge field.  For the
application to multiple M5-branes, it should be identified with
the KK modes.

\end{abstract}

\end{titlepage}

\setcounter{footnote}{0}

\section{Introduction and summary}

The description of multiple M5-branes has been a long-standing
challenging issue for string physicists.  The precise formalism of this system
will enable us to understand non-perturbative properties of string
and gauge dynamics such as duality.  As a supersymmetric theory,
it is supposed to be described by an interacting two-form self-dual gauge field with
its super-partners, and called a $(2,0)$-theory.
Recently, there was a proposal \cite{Douglas:2010iu,Lambert:2010iw}
to describe multiple M5-branes
as non-perturbative dynamics of D4-branes, instead of 6D two-form
gauge fields, with the instanton number corresponding to the KK momentum.
While this is a possible scenario, there are
still various proposals to formulate non-Abelian two-form gauge fields
\cite{r:LP, r:HHM, r:MM5a, r:MM5b, apply}.

In the mathematical literature, the geometry described by a non-Abelian
two-form gauge field is referred to as the non-Abelian gerbe \cite{r:BM}
and has recently been rapidly developed.  A readable review \cite{r:BH}
on this subject was published a few years ago.
There is a general description of two-form gauge fields in terms of
the crossed module. It is natural to apply this general framework
to the $(2,0)$-theory \cite{r:HHM, apply}.

In this note, after a short review of \cite{r:BH}, we explain in \S \ref{s:review}
a constraint on gauge fields which simplifies the gauge transformation.
In \cite{r:BH}, it was emphasized that this constraint,
which demands the so-called ``fake curvature'' to vanish,
is necessary to
define a Wilson surface which describes the parallel transportation of
a string.  Here we point out that it is also natural to introduce such
a constraint in order to define the gauge-invariant action,
as well as defining a gauge symmetry for gauge parameters, which
is an essential feature of the Abelian two-form gauge field.

In \S \ref{s:constraint}, after defining an action which
implement such constraint,
we argue that this constraint restricts the system too tightly.
In particular, we show that the curvature may have
nonvanishing component only in the center of Lie algebra.
For some examples of crossed modules considered in \cite{r:BH},
the physical system associated with such gauge fields
reduces either to a free system or a topological BF-type gauge theory.
In this sense, it seems hard to apply it directly to the $(2,0)$-theory
as it is.

In \S\ref{s:modification}, we propose a modification of
the transformation laws for a special case of crossed modules
which is a generalization of \cite{r:HHM}.
A noticeable feature of this construction is that
the field strength transforms covariantly
without the necessity of imposing any constraint,
as opposed to the situation of a generic non-Abelian gerbe \cite{r:BH}.
This enables us to write an action for such a system without constraints.
It also ensures the existence of a gauge symmetry for the
gauge parameters, which is again a desirable feature.
The action and its equation of motion are analyzed.
We find that in general the two-form gauge potential acquires
mass after a generalized Stueckelberg mechanism.
This may be disappointing to identify them as the massless
two-form gauge potential in the 6D $(2,0)$-theory.
On the other hand, this is natural from the viewpoint of \cite{r:HHM},
where the massive 2-form potential are the KK modes of the two-form potential
due to the compactification on a circle.

Finally, we describe how non-Abelian gerbes can be constructed
geometrically by gluing local patches together via transition functions
on a manifold in \S\ref{geometry}.
This will allow us to discuss topologically nontrivial non-Abelian gerbes.

\section{Non-Abelian gauge symmetry
for two-form gauge \\ potential and constraint
}
\label{s:review}

We review the basic construction of the gauge symmetry for two-form
gauge potential following Ref.\cite{r:BH} and we use the same notation
as much as possible.

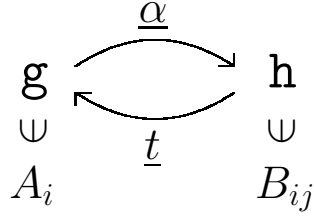
\begin{figure}[bpt]
\setlength{\unitlength}{1pt}
\begin{center}
\vskip 2em
\begin{picture}(90,70)(-45,-40)
\put(-51,-3){\LARGE$\ttg$}
\put(-51,-25){\rotatebox{90}{\Large$\in$}}
\put(-55,-45){\Large$A_i$}
\put(43,-3){\LARGE$\tth$}
\put(43,-25){\rotatebox{90}{\Large$\in$}}
\put(38,-45){\Large$B_{ij}$}
\qbezier(-30,5)(0,25)(30,5)
\put(-6,22){\Large$\ua$}
\put(30,5){\line(-0,1){5}}
\put(30,5){\line(-1,0){6}}
\qbezier(-30,-5)(0,-25)(30,-5)
\put(-4,-30){\Large$\ut$}
\put(-30,-5){\line(0,-1){5}}
\put(-30,-5){\line(1,0){6}}
\end{picture}
\end{center}
\caption{Crossed module}
\label{f:cm}
\end{figure}

Unlike the Abelian case, the non-Abelian gauge
symmetry for two-form
gauge field $B$ is realized with the help
of 1-form gauge field $A$.  While they can take
their values in different Lie algebras, say $A\in \ttg$
($\ttg$ is the Lie algebra for a gauge group $G$)
and $B\in \tth$
($\tth$ is the Lie algebra for a gauge group $H$),
there should be a set of maps (homomorphisms),
$t:H\rightarrow G$ and $\alpha:G\rightarrow \mathrm{Aut}(H)$
which relate these two Lie groups. Here Aut$(H)$ is the automorphism
group of $H$.
(Fig. \ref{f:cm})
These maps are constrained by two conditions
\ba\label{cons}
\alpha(t(h))(h') = h h' h^{-1},\quad
t(\alpha(g) h) =g t(h) g^{-1}
\ea
for any $g\in G$ and $h, h'\in H$.
We denote the Lie algebras of $G$ and $H$ as $\ttg, \tth$.
The differential form of the homomorphisms $t,\alpha$ are written as
$\ut, \ua$, respectively. The consistency conditions
(\ref{cons}) become
\ba\label{cons2}
\ua(\ut(y))(y')=[y,y'],\quad
\ut(\ua(x)y)=[x,\ut(y)]
\ea
for any $x\in \ttg$ and $y,y'\in \tth$.
Since they are homomorphisms, it implies
\ba
\ut([y,y'])=[\ut(y),\ut(y')],\quad
\ua([x,x'])=[\ua(x),\ua(x')],\quad
\ua(x)([y,y'])=[\ua(x)(y),y']+[y,\ua(x)(y')]
\ea
for any $x,x'\in\ttg$ and $y,y'\in \tth$.
The set $(G,H,t,\alpha)$ which satisfies these
constraints is called a crossed module.
While $G$ and $H$ can be arbitrary, the existence of maps $t,\alpha$ gives
a severe constraint on them.
For example, one may wonder if one can define two-form gauge symmetry
without one-form by choosing $G$ to be trivial.  However,
this forces $\ut=0$ and the first consistency condition in
(\ref{cons2}) implies that $H$ is Abelian.


A few examples of crossed modules are following:
\bi
\item[(I)]
$G$ is an arbitrary Lie group, $H=G$, $t$ is the identity map,
and $\alpha(g)(h)=ghg^{-1}$ ($g,h\in G=H$).
\item[(II)]
$G$ is an arbitrary Lie group, $H$ is a vector
space which realizes a representation (say $\rho$)
of $G$ (As a Lie group, $H$ is Abelian),
$\ut(v)=0$, and
$\alpha(g)(v)=\rho(g)v$ ($g\in G$ and $v\in H$).
\item[(III)]
$G$ is a semidirect product $G=K\ltimes V$,
where $K$ is any Lie group, and $V$ is
a vector space which realizes a representation (say $\rho$) of $K$.
Group operation of $G$ is given as $(k,v)\cdot (k', v')=(k\cdot k', v+\rho(k)v')$,
where $(k,v)\in K\ltimes V$.
We also let
$H=V$, $t(v)=(e,v) \in 
G$, where $e$ is the unit element in $K$,
and $\alpha((k,v)(v'))=\rho(k)(v')$.
\ei
The third example is more involved but plays an important role
in section \ref{s:modification}.  Since there are a variety of choices for
crossed modules, we will keep general notation such as $t,\alpha$.
We will refer to these three examples by the labels (I), (II) and (III) below.

We denote the gauge fields on a manifold $\mathcal{M}$ with
local coordinate $x^i$ ($i=1,\cdots,\mbox{dim}{\mathcal{M}}$)
as $A_i(x) dx^i\in \ttg$ and
$B_{ij}(x)dx^i\wedge dx^j \in \tth$.  The gauge transformation
is specified by a function $g(x)\in G$ and a one-form $a_i(x)dx^i \in \tth$ such that
\ba
A'_i&=& gA_i g^{-1}+g\partial_i g^{-1}+\ut(a_i), \label{gt1A}\\
B'_{ij}&=& \alpha(g)(B_{ij}) +\ua(A'_i)(a_j)-\ua(A'_j)(a_i)+\partial_i a_j
-\partial_j a_i +[a_i, a_j]. \label{gt1B}
\ea
The corresponding infinitesimal gauge transformations are
\ba
\delta A_i&= & \partial_i \lambda+[A_i,\lambda]+\ut(a_i), \label{gt2A}\\
\delta B_{ij} &=& -\ua(\lambda)(B_{ij}) +\partial_i a_j
-\partial_j a_i+\ua(A_i) (a_j)-\ua(A_j) (a_i),
\label{gt2B}
\ea
where $\lambda\in \ttg$ and $a_i$ is the infinitesimal version of
the same symbol appearing in (\ref{gt1A},\ref{gt1B}).
This transformation is consistent, that is, the gauge algebra closes,
\ba
\left[\delta_{\lambda^{(1)}, a^{(1)}}, \delta_{\lambda^{(2)}, a^{(2)}}\right]=
\delta_{\lambda^{(3)}, a^{(3)}},
\ea
with
\ba
\lambda^{(3)}&=&[\lambda^{(1)},\lambda^{(2)}],
\\
a^{(3)}_j&=&\ua(\lambda^{(1)})(a^{(2)}_j)
-\ua(\lambda^{(2)})(a^{(1)}_j)\,.
\ea
In the following, for simplicity,
we refer to a geometrical system of gauge potentials $(A_i, B_{ij})$
with such gauge transformations as a ``non-Abelian gerbe",
while this terminology seems to be used in more restrictive sense
in mathematical literature.

There are two types of curvatures
associated with the two gauge fields $A_i$ and $B_{ij}$.
For the one-form $A_i$, we have
an analog of field strength which is called
fake curvature in \cite{r:BH}:
\ba
\cF_{ij}:=\partial_i A_j-\partial_j A_i+[A_i, A_j] -\ut(B_{ij})\in \ttg
\,.
\ea
Under gauge transformation (\ref{gt1A},\ref{gt1B}), it transforms homogeneously,
\ba\label{ftr}
\cF'_{ij}=g \cF_{ij} g^{-1}\,.
\ea
For the two-form $B_{ij}$, the three form field strength
(2-curvature) is defined as
\ba
Z_{ijk}=\sum_{(3)}\left(\partial_i B_{jk}+\ua(A_i)( B_{jk})\right).
\label{defZ1}
\ea
Here $\sum_{(3)}$ represents a sum over cyclic permutations of $(ijk)$
that makes the expression totally anti-symmetric in all indices.
$Z_{ijk}$ transforms as
\ba\label{ztr}
Z'_{ijk}=\alpha(g)( Z_{ijk})+\sum_{(3)}\ua(\cF'_{ij})(a_k)\,.
\ea
The second term breaks the homogeneity of gauge transformation
of $Z_{ijk}$ while the appearance of such term is natural.
It gives rise to an obstruction to gauge invariant
action  in the simplest form
$L\sim\sum_{ijk}\langle Z_{ijk},Z^{ijk}\rangle$.
In order to avoid such obstacle, it may be possible to introduce a constraint
\ba\label{e:vfc}
\cF_{ij}=0
\ea
on $\mathcal{M}$.
This constraint is covariant as
the fake curvature transforms homogeneously.
In \cite{r:BH}, this constraint is motivated
to define the parallel
transport along a surface consistently.


The fake curvature and the 2-curvature satisfy the following identity,
\ba
&&\sum_{(3)} \left(\partial_i \cF_{jk}+[A_i,\cF_{jk}]\right)+\ut(Z_{ijk})=0,
\\
&&\sum_{(4)} \left(\partial_i Z_{jkl}+\ua(A_i) (Z_{jkl})\right)
=\sum_{(6)} \ua(\cF_{ij})(B_{kl}),
\ea
which are the counterparts of the Bianchi identity in ordinary gauge
theories with a one-form gauge potential only.
Here $\sum_{(4)}$ and $\sum_{(6)}$ represent a sum over 4 or 6
terms that totally anti-symmetrize all indices.

\paragraph{Gauge symmetry of gauge parameters
and transformation at triple intersection\\}
As we mentioned, the Abelian two-form gauge field may be obtained
by choosing $G=\{e\}$, $H=U(1)$ and $\ut=\ua=0$,
the one-form gauge potential $A_i$ is identically zero,
and the gauge transformation of the two-form potential is
$\delta B_{ij}=\partial_i a_j-\partial_j a_i$.
In this case, there is a gauge symmetry for the gauge parameter
\ba
\delta a_i=\partial_i\epsilon.
\ea
It reduces the degree of freedom
of the gauge symmetry group.

For non-Abelian case, we expect to have similar redundancy
in the gauge parameters.
For a generic choice for the crossed module,
one may consider the gauge transformation defined by
\ba\label{gg}
\lambda=-\ut(\epsilon),\quad
a_i=\partial_i\epsilon + \ua(A_i)(\epsilon)
\ea
for arbitrary $\epsilon(x) \in \tth$.
The corresponding gauge transformation is
\ba
\delta A_i=0, \quad
\delta B_{ij}=\ua(\cF_{ij})(\epsilon)\,.
\ea
Thus the redundancy exists
only when the fake curvature $\cF$ vanishes.

The redundancy of gauge parameter is a
characteristic feature of two-form gauge fields
and plays a fundamental role in the geometric structure of gerbes.
Consider the base space $\mathcal{M}$ covered by local
coordinate patches $U_\alpha$. On each patch, we have
the gauge fields $A_{(\alpha)}$ and $B_{(\alpha)}$.
On the intersection of two patches,
$U_\alpha\cap U_\beta$, the fields $(A_{(\alpha)}, B_{(\alpha)})$
and $(A_{(\beta)}, B_{(\beta)})$ are matched through
a gauge transformation with parameters $g_{(\alpha\beta)}$, $a_{(\alpha,\beta)}$.
A novel feature of the gerbe which distinguishes itself
from the notion of a bundle is that there may be a
discrepancy of gauge transformations on the triple intersection
$U_\alpha\cap U_\beta\cap U_\gamma$.
More precisely, there is a freedom to
modify the gauge parameter by using the degree of freedom
(\ref{gg}), so that the consistency condition
\ba
g_{(\alpha\beta)}g_{(\beta\gamma)}=g_{(\alpha\gamma)}
\ea
for a bundle is modified to the relation
\ba
g_{(\alpha\beta)}g_{(\beta\gamma)}=t(h_{(\alpha\beta\gamma)})
g_{(\alpha\gamma)}\,.
\ea
Here $h_{(\alpha\beta\gamma)}\in H$ is a finite gauge transformation
that corresponds to the infinitesimal parameter $\epsilon\in\tth$
in (\ref{gg}).  There is an analogous relation for the transformation
parameter $a_i^{(**)}$'s, which is the integrated
version of the second transformation in (\ref{gg}).
Its explicit form is more complicated \cite{r:BH}.
We will come back to this issue below in \S \ref{geometry}.

\section{Implications of vanishing fake curvature condition}
\label{s:constraint}

One may define a gauge invariant action for
the gauge fields $(A_i,B_{ij})$ by noting that
the 2-curvature $Z$ transforms covariantly when $\cF=0$:
\ba \label{lag1}
L=-\frac{1}{12}(Z_{ijk}, Z^{ijk})- \frac{1}{4}
\langle E_{ij},\cF^{ij}\rangle\,,
\ea
where $E_{ij}\in \ttg$ is an auxiliary two-form field.
Variation of $E_{ij}$ produces the condition $\cF_{ij}=0$.
We use $(\cdot, \cdot)$ to denote the inner product in $\tth$
which is invariant under transformations by $G$,
i.e., $(\alpha(g)(A), \alpha(g)(B))=(A,B)$.
Similarly, $\langle\cdot, \cdot\rangle$ is the inner product in $\ttg$ which
is invariant under the adjoint action of $G$.

This action in general has on-shell gauge-invariance
if we define the transformation of $E_{ij}$ as,
\ba
\delta E_{ij}=[E_{ij}, \lambda],
\ea
together with the transformations (\ref{gt2A}, \ref{gt2B}).
While the inhomogeneous term in (\ref{ztr})
breaks the off-shell gauge symmetry,
the equation of motion by the variation of
the auxiliary field $E_{ij}$ imposes the vanishing of
fake curvature condition and recovers the gauge symmetry.

We note that, for some choice of crossed module,
it may be possible to recover the off-shell gauge invariance.
Namely, if there exist a skew bilinear map $\beta:\tth\times\tth \rightarrow \ttg$
such that,
\ba
\beta(y,y')=-\beta(y',y),\quad
\langle \beta(y,y'),x\rangle=(y,\ua(x)(y')),
\ea
one may redefine the gauge transformation of $E_{ij}$ as,
\ba
\delta E_{ij}=[E_{ij},\lambda]-2\beta(Z_{ijk},a^k).
\ea
With this redefinition, the action (\ref{lag1}) has
off-shell gauge invariance.

In the examples of crossed module mentioned in section \ref{s:review},
(I) has such a map, namely,
\ba
\beta(y,y')=[y,y'],\quad y,y'\in \tth\,(=\ttg)\,.
\ea
Similarly, for the example (II) when $H$ is the adjoint representation of
$G$, there exists a similar map.  For such cases, (\ref{lag1})
has off-shell gauge symmetry.

One may also define an action for the self-dual
gauge field $Z=*Z$ in a Lorentz non-covariant manner
 \cite{r:PS, r:HHM}, for example, by taking the
Lorentz non-covariant Lagrangian
\ba
L=-\frac14 ((*Z)_{ab5},(Z^{ab5}+\frac{1}{6}\epsilon^{abcde} Z_{cde}))-\frac14
\langle E_{ij},\cF^{ij}\rangle\,,
\ea
or its possibly Lorentz covariant version \cite{r:PST}.
Here the indices $a,b,c\cdots$ runs from $0$ to $4$.
Except for the appearance of an extra gauge field $A_i$,
which will be eliminated in the constraint in the end,
this seems to give a suitable candidate of
Lagrangian of the non-Abelian two-form gauge fields.

In this paper, we will not pursue detailed properties of such action
further but focus on the implication of
the vanishing of the fake curvature condition:
\ba
\cF_{ij}=F_{ij}-\ut(B_{ij})=0,\quad
F_{ij}:=\partial_i A_j-\partial_j A_i+[A_i, A_j]\,,
\label{cF=0}
\ea
since it imposes a strong constraint on $B$
which seems problematic to be applied to M5.
Taking a covariant derivative of the constraint
and using Bianchi identity, we obtain
\ba\label{bianchi}
0=\sum_{(3)}(\partial_iF_{jk}+[A_i,F_{jk}])=\ut(Z_{ijk}).
\ea
It implies that the 2-curvature can only take values in elements
$y\in \tth$ with $\ut(y)=0$.
From (\ref{cons2}), such an elements should
commute with all the element in $\tth$,
\ba
0=\ua(\ut(y))(y')=[y,y']\,.
\ea
Hence nonvanishing components of $Z$ belong to the center of $\tth$.
It is disappointing that,
as we will see in the concrete examples
mentioned in section \ref{s:review}, the system
seems to become either free or topological.

Let us now consider each example of crossed modules in \S \ref{s:review}
in some details.
\bi
\item[(I)]
Since $\ut$ does not have nontrivial kernel, (\ref{bianchi}) implies that
$Z=0$.  Thus the first term in (\ref{lag1}) vanishes after
imposing the constraint. The Lagrangian becomes $L=\langle E,\cF\rangle$,
which may be regarded as an analogue of the topological BF theory.

\item[(II)]
Since $H$ is Abelian and $\ut=0$, the constraint $\cF_{ij}=0$ implies $F_{ij}=0$.
Thus $A_i$ should be a pure gauge and can be absorbed by the gauge
transformation. After taking $A_i=0$ the action reduces to
that of a set of Abelian two-form potentials since $Z=dB$.
Interaction through the non-Abelian group $G$ is lost,
and the theory is free.

\item[(III)]
Since $t: V\rightarrow K\ltimes V$ ($t(v)=(e,v)$) is one-to-one,
$Z_{ijk}=0$. So the Lagrangian (\ref{lag1}) reduces to BF type theory.
\ei
In these examples considered so far,
the constraint (\ref{cF=0})
gives either a free or topological theory.
In either case, it does not fit with
the purpose of application to M-theory.

Let us give a brief (and incomplete)
argument that the situation does
not improve in general. As we mentioned,
the kernel of $t$ can be identified
with the center $C$ of $H$.
Let $T^a$ ($a=1,\cdots,n$) be the generators in $C$ and
$T^\alpha$ the rest of the generators in $H$. The structure constant has the
form $[T^a, T^P]={f^{aP}}_Q T^Q=0$, where $P,Q$ run over all generators of $\tth$
since $T^a$ is in the center.  Suppose that $H$ has
a positive definite metric and $f^{PQR}$ be anti-symmetric
in $PQR$.  It implies that
${f^{PQ}}_a=0$.
As a result,
\ba
&Z_a=(DB)_a=dB_a +{f^{PQ}}_a A_P B_Q= dB_a,
\\
&Z_\alpha = 0.
\ea
Hence the 2-cuvature becomes either free (namely $Z=dB$) or
vanishes.  In this rough argument, it seems reasonable
to suspect that a system with vanishing fake curvature
is either free or topological.

While it may be still possible to have a system with interaction
if the metric of $H$ is not positive definite,
we try to find a possible modification of the non-Abelian
gauge symmetry where we do not need the condition (\ref{cF=0}).

\section{A modification of non-Abelian gerbe}
\label{s:modification}

\subsection{Definition}

The origin of the condition (\ref{e:vfc}) is
the inhomogeneity of gauge transformation for the 2-curvature.
A hint to obtain homogeneous transformation for 2-connection
can be found in \cite{r:HHM}.
Here we give a generalization of \cite{r:HHM}
which seems to suggest a possible modification of
the definition of non-Abelian gerbes.



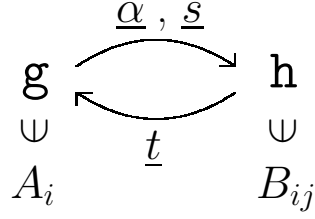
\begin{figure}[bpt]
\setlength{\unitlength}{1pt}
\begin{center}
\vskip 2em
\begin{picture}(90,70)(-45,-40)
\put(-51,-3){\LARGE$\ttg$}
\put(-51,-25){\rotatebox{90}{\Large$\in$}}
\put(-55,-45){\Large$A_i$}
\put(43,-3){\LARGE$\tth$}
\put(43,-25){\rotatebox{90}{\Large$\in$}}
\put(38,-45){\Large$B_{ij}$}
\qbezier(-30,5)(0,25)(30,5)
\put(-15,22){\Large$\ua$\;, $\us$}
\put(30,5){\line(-0,1){5}}
\put(30,5){\line(-1,0){6}}
\qbezier(-30,-5)(0,-25)(30,-5)
\put(-4,-30){\Large$\ut$}
\put(-30,-5){\line(0,-1){5}}
\put(-30,-5){\line(1,0){6}}
\end{picture}
\end{center}
\caption{Crossed module with an extra map $s$}
\label{f:cm2}
\end{figure}

We start with the information of a crossed module $(G, H, \alpha, t)$,
but with the addition of a map $s$ (see Fig.\ref{f:cm2}),
\ba
\alpha&:& G \rightarrow \mathrm{Aut}(H), \\
t&:& H \rightarrow G, \\
s&:& G \rightarrow H,
\ea
so that their differential forms satisfy
the following conditions
\ba
&&\ua(\ut(y))(y') = [y, y'],
\label{c1} \\
&&\ut(\ua(x)(y)) = [x, \ut(y)],
\label{c2} \\
&&\ua([x, \ut\cdot\us(x')])(\us(x'')) = 0,
\label{c3} \\
&&\ua(x)(\us(x'))-\ua(x')(\us(x)) = \us([x, x']),
\label{c4} \\
&&\ua(x)((1-\us\cdot\ut)(y)) = 0
\label{c5}
\ea
for all $x, x', x'' \in \ttg$ and $y, y' \in \tth$.
We have denoted both Lie algebras of $G, H$ as $\ttg, \tth$,
respectively,
and the Lie brackets for both $\ttg$ and $\tth$ as $[\cdot, \cdot]$.
The first two conditions are identical to (\ref{cons2}).
Last three conditions are new and necessary
to make the modified gauge transformations
to be closed.  We also need to demand
the group $H$ to be Abelian.

The one-form potential $\bA_i \in \ttg$ and
the two-form potential $\tilde{B}_{ij} \in \tth$ transform as
\ba
\delta \bA_i &=&
\partial_i \bL + [\bA_i, \Lambda] + \ut(\tilde{a}_i),
\label{deltaA-1} \\
\delta \tilde{B}_{ij} &=&
\partial_i\tilde{a}_j - \partial_j\tilde{a}_i + \ua(\bA_i)(\tilde{a}_j) - \ua(\bA_j)(\tilde{a}_i)
- \ua(\bL)(\tilde{B}_{ij}) + \ua(\bF_{ij})(\us(\bL)),
\label{deltaB-1}
\ea
with the gauge transformation parameters
$\bL \in \ttg$ and $\tilde{a}_i \in \tth$.
The last term in (\ref{deltaB-1}) is the only modification to
the gauge transformation laws we made to
the theory of non-Abelian gerbes reviewed in \S \ref{s:review}.
Here we use bold face letters for objects in $\ttg$,
and letters with tildes for objects in $\tth$.

The gauge transformation algebra for (\ref{deltaA-1}) and (\ref{deltaB-1}) is closed
\ba
[\delta_{(1)}, \delta_{(2)}] = \delta_{(3)},
\ea
and the composition rule for the gauge transformation parameters is given by
\ba
\bL_{(3)} &=& [\bL_{(1)}, \bL_{(2)}], \\
\tilde{a}_{(3)i} &=&
\ua(\bL_{(1)})(\tilde{a}_{(2)i})
- \ua(\bL_{(2)})(\tilde{a}_{(1)i}).
\ea

The curvatures are defined by
\ba
\bF_{ij} &=&
\partial_i \bA_j - \partial_j \bA_i + [\bA_i, \bA_j] - \ut(\tilde{B})_{ij},
\label{def-F-1}
\\
\tilde{Z}_{ijk} &=&
\sum_{(3)}\left(\partial_i \tilde{B}_{jk} + \ua(\bA_i)(\tilde{B}_{jk})
- \ua(\bF_{ij})(\us(\bA_k))\right)\,.
\label{def-Z-1}
\ea
Note that the last term in (\ref{def-Z-1}) is added
in comparison with its definition (\ref{defZ1}) in \S \ref{s:review}.
They transform as
\ba
\delta \bF_{ij} &=&
[\bF_{ij}, (1-\ut\cdot\us)(\bL)],
\label{transf-F}
\\
\delta \tilde{Z}_{ijk} &=&
- \ua((1-\ut\cdot\us)\bL)(\tilde{Z}_{ijk}).
\label{transf-Z}
\ea
As opposed to the case in \S \ref{s:review},
the covariance of the field strength $\tilde{Z}_{ijk}$ does not
rely on the vanishing of the fake curvature $\bF_{ij}$ here.
The operator $(1-\ut\cdot\us)$ plays a role of projection
in $\ttg$ such that $(1-\ut\cdot\us)\ttg$ becomes a subalgebra of $\ttg$.

Furthermore, unlike the set-up in \S \ref{s:review},
there is always a gauge symmetry of gauge parameters for arbitrary $\bF$.
Consider a variation of gauge parameters,
\ba
\delta \tilde a_i=\partial_i \tilde{\kappa} +\ua(\bA_i) (\tilde{\kappa}),
\qquad
\delta \bL=-\ut(\tilde{\kappa})
\label{kappa}
\ea
for arbitrary $\tilde{\kappa}\in \tth$.
The gauge transformation is invariant under this variation.
As we mentioned earlier,
such a symmetry is used to allow a relaxation of the consistency condition
of transition functions
at the intersection of three patches.

The Bianchi identities are
\ba
&\sum_{(3)}\left(\partial_i\bF_{jk}+[\bA_i,\bF_{jk}]\right)+\ut(\tZ_{ijk})=0,\\
&\sum_{(4)}\left(\partial_i\tZ_{jkl}+\ua(\bA_i)(\tZ_{jkl})\right)+\sum_{(6)}\ua(\bF_{ij})
(\us(\bF_{kl}))=0\,.
\ea


\subsection{Explicit construction of $(G, H, \alpha, t, s)$}
\label{s:explicit}

A class of nontrivial examples of the data $(G, H, \alpha, t, s)$
needed in our definition of a non-Abelian gerbe is given as follows.
\footnote{
The same construction is given in a different but equivalent description
in the appendix.
The formulae here are more explicit,
while those in the appendix are more compact.
}
We start with the example (III) of crossed modules in \S \ref{s:review}. Namely,
we define $G$ as the semidirect product $G=K\ltimes V$
with the group multiplication rule,
\ba
(k_1, v_1)\cdot (k_2, v_2)=(k_1\cdot k_2,
v_1+\rho(k_1)v_2)\,,
\quad k_1, k_2\in K, \quad
v_1, v_2\in V\,.
\ea
An element in $\ttg$ can be denoted as
$(x,y)\in \ttg$ with $x\in \ttk$ and $y\in V$.  The commutator of $\ttg$
is defined as
\ba
[(x_1, y_1), (x_2, y_2)]=([x_1,x_2], \ur(x_1)(y_2)-\ur(x_2)(y_1))\,.
\ea
We identify $\tth$ with $V$, and let
\ba
\ut(y) &:=&(0, M y)\qquad y\in V, \\
\ua(x,y_1)(y_2)&:=& \ur(x)(y_2)\qquad
x\in \ttk,\, y_1,y_2\in V\,,\\
\us(x,y)&=&M^{-1} y,
\ea
where $M$ is an invertible linear map
$M$ on $V$ with the property
\ba
[M,\ur(x)]=0, \qquad \forall x\in \ttk\,.
\ea
By Schur's lemma, $M$ should proportional to identity
matrix in the irreducible subspace of $V$.

Loosely speaking $\us$ is the inverse of the map $\ut$ and satisfies
\ba
\us\cdot \ut(y)=y,\qquad
\ut\cdot\us(x,y)=(0,y)
\ea
for arbitrary $x\in \ttk$ and $y\in V$.
Note that the map $\ut\cdot\us$ is a projection from $\ttg$ to $\tth$.
We will also need its complementary projection
\ba
\Pi := 1 - \ut\cdot\us,
\label{Pi}
\ea
which projects $\ttg$ to $\ttk$ as
\ba
\Pi(x, y) = (x, 0).
\ea
One can check that all the requirements (\ref{c1})--(\ref{c5})
are satisfied.

We introduce two one-form gauge potentials
$A_i\in \ttk$ and $\tilde A_i\in V$,
so that $\bA_i:=(A_i,\tilde A_i)\in \ttg$,
and a two-form gauge potential $\tilde{B}_{ij}\in V$.
Associated with them, we introduce zero-form gauge parameters
$\Lambda\in \ttk$, $\tilde\Lambda\in V$,
so that $\bL:=(\Lambda,\tL)\in \ttg$, and $\tilde a_i\in V$.

The gauge transformations (\ref{deltaA-1}) and (\ref{deltaB-1}) can be matched with
(\ref{gt2A}) and (\ref{gt2B}) up to the last term in (\ref{deltaB-1}).
Explicitly,
\ba
\delta A_i&=& \partial_i \Lambda+[A_i,\Lambda],
\label{ddA1}\\
\delta \tilde A_i &=& \partial_i\tilde\Lambda +\ur(A_i)(\tilde\Lambda)
-\ur(\Lambda)(\tA_i)+M\ta_i,
\label{ddA2}\\
\delta \tB_{ij}&=& \partial_i\ta_j-\partial_j\ta_i+\ur(A_i)(\ta_j)-\ur(A_j)(\ta_i)
-\ur(\Lambda)(\tB_{ij})+M^{-1}\ur(F_{ij})(\tL)\,.
\label{ddB}
\ea
According to (\ref{def-F-1}) and (\ref{def-Z-1}),
the field strength $F_{ij}\in \ttk$ and
$\tF_{ij}, \tZ_{ijk}\in V$ are
\ba
F_{ij}&=& \partial_i A_j-\partial_j A_i+[A_i, A_j],
\\
\tF_{ij}&=& \partial_i \tA_j-\partial_j\tA_i+\ur(A_i)(\tA_j)-\ur(A_j)(\tA_i)-M\tB_{ij},
\label{ddtF}\\
\tZ_{ijk}&=& \sum_{(3)}\left(\partial_i \tB_{jk}+\ur(A_i)\tB_{jk}-M^{-1} \ur(F_{ij})(\tA_k)\right),
\label{defZ2}
\ea
such that $\bF_{ij}=(F_{ij}, \tF_{ij})\in \ttg$.
Following (\ref{transf-F}) and (\ref{transf-Z}),
all curvatures transform homogeneously,
\ba
\delta F_{ij}&=& [F_{ij},\Lambda]\,,\\
\delta \tF_{ij}&=& -\ur(\Lambda)(\tF_{ij})\,,\\
\delta \tZ_{ijk}&=& -\ur(\Lambda)(\tZ_{ijk})\,.
\ea
The subalgebra
$(1-\ut\cdot\us)\,\ttg$ is identified with $\ttk$.



The Bianchi identity for curvatures are
\ba
&&\sum_{(3)}(\partial_i F_{jk}+[A_i, F_{jk}])=0, \\
&&\sum_{(3)}(\partial_i \tF_{jk}+\ur(A_i)(\tF_{jk}))+M\tZ_{ijk}=0, \\
&&\sum_{(4)}\left(\partial_i \tZ_{jkl}+\ur(A_i)(\tZ_{jkl})\right)+
\sum_{(6)}\ur(F_{ij})(M^{-1} \tF_{kl})=0\,.
\ea

The dependence of the gauge algebra on $M$ can be undone
through the redefinition
\ba
\tA_i = M\tA'_i, \qquad
\tF_i = M\tF'_i, \qquad
\tilde\Lambda = M \tilde\Lambda'.
\label{redef-M}
\ea
In terms of $\tA', \tF', \tilde\Lambda'$,
the map $M$ disappears from all equations above.

We suspect that the data $(G,H,\alpha, t,s)$ which satisfies (\ref{c1}--\ref{c5})
may be restricted to this example while we do not have a general proof.

\subsection{Lagrangian formalism}

Since the gauge transformation is homogeneous for all curvatures,
the construction of gauge invariant action is straightforward.
For simplicity, we use the modified non-Abelian gerbe
in the previous subsection.
A standard gauge invariant action is,
\ba\label{LFtFtZ}
L=-\frac14 \Tr(F_{ij})^2-\frac14 \langle\tF_{ij},\tF^{ij}\rangle
-\frac{1}{12}\langle\tZ_{ijk},\tZ^{ijk}\rangle\,.
\ea
Here $\langle\cdot,\cdot\rangle$ is an inner product
in $V$ which is invariant under the action of $K$.
We notice that this system has a natural structure where
the gauge field $\tB_{ij}$ acquire mass through
Stueckelberg mechanism.

In order to see it, we take a look at the gauge transformation of
$\tA_i$ (\ref{ddA2}).  It is possible to take
$\tA_i=0$ by choosing $\ta_i$.  With this gauge fixing,
$\tF_{ij}$ (\ref{ddtF}) becomes $-M\tB_{ij}$.
By putting this value back to (\ref{LFtFtZ}), one obtains,
\ba\label{LFtFtZa}
L=-\frac14 \Tr(F_{ij})^2
-\frac{1}{12}\langle\tZ_{ijk},\tZ^{ijk}\rangle
-\frac14 \langle M\tB_{ij}, M\tB^{ij}\rangle\,.
\ea
The matrix $M$ is proportional to unit matrix
for each irreducible representation.  We decompose
the vector space $V$ into irreducible subspaces,
\ba
V=\oplus_\sigma V^{(\sigma)}
\ea
then $M$ is written as,
\ba
M=\oplus_\sigma m^{(\sigma)} E^{(\sigma)}
\ea
where $E^{(\alpha)}$ is the unit matrix acting on $V^{(\alpha)}$.
If we decompose $\tB_{ij}$ into components $\tB_{ij}^{(\sigma)}$,
the action (\ref{LFtFtZa}) gives mass $|m^{(\sigma)}|$ for $\tB_{ij}^{(\sigma)}$.


Roughly speaking, $\tA_i$ plays the role of the Stueckelberg field 
with respect to $\tB_{ij}$, 
and in a perturbative theory we expect that
$\tB_{ij}$ eats up $\tA_i$ to acquires mass.
This is a generic feature of our formulation of non-Abelian gerbes.

Interestingly, the interaction remains for $\tB_{ij}^{(\sigma)}$
through the gauge field $A_i$ even after the mass generation.
This is a novel feature
which was not present in the examples of \S\ref{s:constraint}.
There is no direct interaction among $\tB_{ij}^{(\sigma)}$ with different
representations while there may be a mixing among fields
in an identical representation.

If the base manifold $\mathcal{M}$ has five dimensions, 
one may impose a duality condition
between $\tA$ and $\tB$ as,
\ba\label{sdeq}
\tZ_{ijk}^{(\sigma)}=\frac12 \epsilon_{ijklm}\tF^{(\sigma)lm}\,,
\ea
since these fields transform the same way.
Again, at the linear order, one may replace the right hand side by
$\frac{m^{(\sigma)}}2 \epsilon_{ijklm}\tB^{(\sigma)lm}$
which gives rise to mass $|m^{(\sigma)}|$ for $\tB^{(\sigma)}_{ij}$.
We note that this is
a generalization of the non-Abelian self-dual gauge theory 
for multiple M5-branes compactified on a circle \cite{r:HHM}. 
More precisely,
on the uncompactified large 5 dimensional worldvolume,
the gauge theory on multiple M5-branes is an example 
of the non-Abelian gerbes defined in this paper.
There $V$ is replaced by
the infinite tensor product of adjoint representations
for KK tower  and $|m^{(\sigma)}|$ 
by KK mass.  The gauge invariant action which produces
(\ref{sdeq}) was also constructed.  In this set-up,
$\tA_i$ and $\tB_{ij}$ are identified with KK modes of
$B_{i5}$ and $B_{ij}$ in six dimensions
while $A_i$ be identified with the zero mode of
$B_{i5}$ and $K$ be identified with $SU(N)$ 
with $N$ the number of M5 branes.
While we do not repeat the arguments here, 
we believe that this is the most natural set-up
for non-Abelian gerbe for the description of the multiple M5-branes.


Instead of using infinite tensor product of
adjoint representations for $V$, we may use
arbitrary representation of $K$.
This may be helpful to understand the $O(N^3)$ entropy
issue of multiple M5 branes through changing the
representations for KK modes (see for example \cite{Kim:2011mv}).

\section{Geometric construction of non-Abelian gerbes}
\label{geometry}

\subsection{Finite gauge transformation}

For the convenience of the discussions later,
we need a description of the finite gauge transformations,
which can be viewed as 
an exponentiation of the infinitesimal gauge transformations 
introduced above.
For clarity, we use the explicit form of $(G,H,\alpha,t,s)$
given in subection \ref{s:explicit}. 
Namely, we have $G=K\ltimes V$ and $H=V$, etc. 
But the formulae here can be trivially extended to the general
case of $(G,H,\alpha,t,s)$.
Without loss of generality, we set $M=1$ in the definition of $\ut, \us$.

The finite gauge transformation parameters are
\ba
g=(k,\tL)\in K\ltimes V,\quad
\tilde{a}_i\in V\,.
\ea
As we defined, the composition of two elements in $K\ltimes V$ is
$
(k_1,\tL_1)\cdot(k_2,\tL_2)=(k_1 k_2, \tL_1+\rho(k_1)\tL_2)\,,
$
so the inverse of $g=(k,\tL)$ is
$
g^{-1}=(k^{-1},-\rho(k^{-1})(\tL))\,.
$
The adjoint action of $g$ on 
$\bA=(A,\tA)\in \ttg$ is
\ba
g\cdot \bA \cdot g^{-1}=(k,\tL)(A,\tA)(k^{-1},-\rho(k^{-1})(\tL))
=(k A k^{-1}, \rho(k)(\tA)-\ur(kAk^{-1})(\tL))\,.
\ea
With the help of these formulae, the gauge transformation
(\ref{gt1A}) becomes
\ba
\bA'_i&=& (A'_i, \tA'_i)=g\bA_i g^{-1}+g\partial_i g^{-1}+\ut(\tilde{a}_i)\,,
\\
A'_i&=& k A_i k^{-1}+k\partial_i k^{-1}\,,
\label{gt3A}\\
\tA'_i &=& \rho(k)(\tA_i) -(\partial_i+\ur(A_i'))(\tL)+\tilde{a}_i\,.
\label{gt3A'}
\ea
The gauge transformation (\ref{gt1B}) is modified to
\ba
\tB'_{ij}&=&\alpha(g)(\tB_{ij}) +\ua(\bA_i')(\ta_j)-\ua(\bA_j')(\ta_i)
+\partial_i \ta_j-\partial_j \ta_i
+ g\us(\bF_{ij})g^{-1} -\us(g \bF_{ij} g^{-1})\nn\\
&=& \rho(k)(\tB_{ij}) +\ur(A_i')(\ta_j)-\ur(A_j')(\ta_i)
+\partial_i\ta_j-\partial_j\ta_i
+\ur(k F_{ij}k^{-1}) (\tL)\,.
\label{gt3B}
\ea
These transformation laws are consistent with
the infinitesimal gauge transformations (\ref{ddA1}--\ref{ddB}).
They also constitute a closed algebra
of gauge transformations.  Define
the operator $T(k, \tL, \ta)$ as the gauge transformation 
with gauge parameters $(k, \tL)\in G$ and $\ta_i\in H$.
One may show that,
\ba
&&T(k^{(1)}, \tL^{(1)}, \ta^{(1)}_i)\cdot
T(k^{(2)}, \tL^{(2)}, \ta^{(2)}_i)(\Phi)\nn\\
&&~~~~~~~~~~
=T(k^{(1)}k^{(2)}, \tL^{(1)}+\rho(k^{(1)})(\tL^{(2)}), \ta^{(1)}_i
+\rho(k^{(1)})(\ta^{(2)}_i))(\Phi)
\label{TTT}
\ea
for $\Phi=A_i, \tA_i, \tB_{ij}$.

Let us denote the parameter space of gauge transformations by
\ba
{\cal G} := \{ (k, \tL, \ta) \in (G, V) \}.
\label{calG}
\ea
The composition rule (\ref{TTT}) of gauge transformations induces 
the notion of a product in ${\cal G}$
\ba
(k^{(1)}, \tL^{(1)}, \ta^{(1)}_i)\cdot
(k^{(2)}, \tL^{(2)}, \ta^{(2)}_i)
= (k^{(1)}k^{(2)}, \tL^{(1)}+\rho(k^{(1)})(\tL^{(2)}), \ta^{(1)}_i
+\rho(k^{(1)})(\ta^{(2)}_i)).
\ea

The redundancy of the gauge transformation 
can be written as (with $g=(k,\tL)\in G$ and 
$\tO\in V$)
\ba\label{lred}
 \qquad (k',\tL') =(k,\tL+\tO),\quad
\ta_i'=\ta_i +\partial_i \tO+\ur(A_i')(\tO)\,.
\ea
The first transformation is the component form of $g'=t(\tO)g$.
Correspondingly,
we define an automorphism $R(\tO)$ on the parameter space 
${\cal G}$ of gauge transformations by
\ba
R(\tO)(k, \tL, \ta) = (k', \tL', \ta')
\label{RtO}
\ea 
so that 
\ba
T(R(\tO)(k, \tL, \ta)) = T(k, \tL, \ta)
\qquad \forall \; \tO \in V.
\ea
From (\ref{lred}) we see that $R(\tO)$ depends on
the potential $A'$.

The finite gauge transformations of
the field strengths are
\ba
F_{ij}'=k F_{ij}k^{-1},\quad
\tF_{ij}= \rho(k)(\tF_{ij}),\quad
\tZ_{ijk}= \rho(k)(\tZ_{ijk})\,.
\ea

\subsection{Non-Abelian gerbes as a generalization of bundles}

The notion of nontrivial bundles is important in theoretical physics.
A nontrivial bundle, e.g. that of a magnetic monopole,
can be constructed
by specifying transition functions on a manifold.
By analogy, we expect the existence of topologically nontrivial (non-)Abelian gerbes
that can be constructed by patching local descriptions of gauge potentials
together via transition functions.
While the discussions above are sufficient for describing local fluctuations,
we also want to understand the topological issues of non-Abelian gerbes.
In this section we aim to understand the geometrical data
for constructing a non-Abelian gerbe from local patches.
Despite differences in the rules of gauge transformations,
the basic idea behind the construction is essentially the same 
as that of other definitions of gerbes in the literature.
(See e.g. Sec. 5.3 of \cite{r:BH}.)

A fiber bundle defined on a manifold ${\cal M}$
covered by local coordinate patches $U_{\alpha}$
can be constructed by specifying
transition functions $g_{(\alpha\beta)}$
belonging to the structure group $G$
on each intersection
of two patches $U_{\alpha\beta} := U_{\alpha}\cap U_{\beta}$.
Over the intersection of three patches
$U_{\alpha\beta\gamma} := U_{\alpha}\cap U_{\beta}\cap U_{\gamma}$,
the consistency condition for the transition functions is
\ba
g^{(\alpha\gamma)} = g^{(\alpha\beta)}g^{(\beta\gamma)}.
\label{old-consistency}
\ea
This is essentially the rule of composition of gauge transformations.

Similarly, the gauge transformation (\ref{gt3A})--(\ref{gt3B}) defined above
allows us to patch together locally defined potentials $\bA, \tilde{B}$
on a manifold ${\cal M}$.
On the overlap of two local coordinate patches $U_{(\alpha\beta)}$,
we match the potentials on each patch via transition functions
$(k^{(\alpha\beta)}, \tL^{(\alpha\beta)},\tilde{a}^{(\alpha\beta)}_i)$ as
\ba
T^{(\alpha\beta)}(\Phi^{(\beta)})=\Phi^{(\alpha)},\quad
T^{(\alpha\beta)}:=T(k^{(\alpha\beta)}, \tL^{(\alpha\beta)},\tilde{a}^{(\alpha\beta)}_i)
\label{Bg}
\ea
for $\Phi=A_i, \tA_i, \tB_{ij}$.
Consistency of the local patches imply that
\ba
T^{(\alpha\gamma)} =
T^{(\alpha\beta)}T^{(\beta\gamma)}.
\ea

A crucial difference between the ordinary gauge symmetry of
1-form potential and the gauge symmetry of gerbes is
that the latter has a degeneracy in parametrizing gauge transformations
(\ref{lred}).  
This equivalence relation is generated by $R(\tO)$ (\ref{RtO})
as an automorphism on the parameter space ${\cal G}$ of gauge transformations.
It leads us to generalize the notion of fiber bundles by
relaxing this consistency condition (\ref{old-consistency}) to
\ba
p^{(\alpha\gamma)} &=&
R^{(\alpha\beta\gamma)}(p^{(\alpha\beta)}\cdot p^{(\beta\gamma)}),
\qquad
R^{(\alpha\beta\gamma)}:=R(\tO^{(\alpha\beta\gamma)})
\label{new-consistency1}
\ea
for some $\tO^{(\alpha\beta\gamma)} \in V$.
Here we used $p^{(\alpha\beta)}$ etc. to denote an element in ${\cal G}$:
$p^{(\alpha\beta)} = (k^{(\alpha\beta)}, \tL^{(\alpha\beta)}, \ta^{(\alpha\beta)})$.
Namely,
\ba
k^{(\alpha\gamma)}&=& k^{(\alpha\beta)} k^{(\beta\gamma)},
\label{kkk} \\
\tL^{(\alpha\gamma)}&=& \tL^{(\alpha\beta)}+\rho(k^{(\alpha\beta)})(\tL^{(\alpha\beta)})
+\tO^{(\alpha\beta\gamma)},
\label{LLLO} \\
\ta^{(\alpha\gamma)}_i&=& \ta^{(\alpha\beta)}_i+\rho(k^{(\alpha\beta)})(\ta^{(\beta\gamma)}_i)
+ \partial_i \tO^{(\alpha\beta\gamma)}
+\ur(A^{(\alpha)}_i)(\tO^{(\alpha\beta\gamma)}).
\label{new-consistency2}
\ea
These are essentially the composition laws of two consecutive
gauge transformations twisted 
by a transformation (\ref{lred}) of the gauge symmetry
of gauge symmetry.
While the composition for gauge parameters $k,\tL$
depends only on the twist parameter $\tO^{(\alpha\beta\gamma)}$,
the gauge field $A^{(\alpha)}_i$ appears in the transformation
of $\ta^{(\alpha\gamma)}_i$.  In the Abelian case, such a term
does not show up, and it is a new feature of non-Abelian gerbes.
This difference is one of the reasons that make non-Abelian gerbes 
less straightforward to define than its Abelian counterpart.




Eq.(\ref{new-consistency1}) is at the same time
a consistency condition
and a definition of $\tO^{(\alpha\beta\gamma)}$
\footnote{
Here we assume that there is no redundancy in the parametrization of
the gauge symmetry of gauge symmetry by $\tO \in V$.
}
associated to each intersection $U_{(\alpha\beta\gamma)}$
of 3 patches,
as part of the data for the geometric construction of a gerbe.
To proceed further, we have to be more explicit 
and divide the information in $p^{(\alpha\beta)}$ 
into two parts $(g^{(\alpha\beta)}, \ta^{(\alpha\beta)})$,
as they have to be treated differently.
For the part of $g^{(\alpha\beta)}$ in $p^{(\alpha\beta)}$,
the equations (\ref{kkk}) and (\ref{LLLO}) can be summarized as
\ba
h^{(\alpha\beta\gamma)} =
g^{(\alpha\beta)}g^{(\beta\gamma)}{g^{(\alpha\gamma)}}^{-1}
\in t(H).
\label{gghg}
\ea
where $h^{(\alpha\beta\gamma)} = (1_k, \tO^{(\alpha\beta\gamma)})$
and $1_k$ represents the unit element in $\ttk$.

As a result of the definition of $h^{(\alpha\beta\gamma)}$ by (\ref{gghg}),
an additional consistency condition for the data
involving $h^{(\alpha\beta\gamma)}$ should
be satisfied on each intersection of 4 patches
$U_{\alpha\beta\gamma\delta} := U_{\alpha}\cap U_{\beta}\cap U_{\gamma}\cap U_{\delta}$
as
\ba
{h^{(\alpha\gamma\delta)}}^{-1} {h^{(\alpha\beta\gamma)}}^{-1}
{h^{(\beta\gamma\delta)}}^{g_{(\alpha\beta)}}
h^{(\alpha\beta\delta)} = 1_G,
\label{consistency-1}
\ea
where
\ba
{h^{(\beta\gamma\delta)}}^{g^{(\alpha\beta)}} :=
g^{(\alpha\beta)} h^{(\beta\gamma\delta)} {g^{(\alpha\beta)}}^{-1},
\ea
and $1_G$ represents the unit element in $G$.

While the relation (\ref{consistency-1}) is derived from (\ref{LLLO})
as a consistency condition on $h^{(\alpha\beta\gamma)}$,
an analogous consistency condition can be derived from (\ref{new-consistency2}).
In terms of 
\ba
\Delta^{(\alpha\beta\gamma)}_i :=
\partial_i \tO^{(\alpha\beta\gamma)}
+\ur(A^{(\alpha)}_i)(\tO^{(\alpha\beta\gamma)}),
\label{DeltaD}
\ea
the consistency condition is
\ba
&h^{(\alpha\beta\delta)}{}^{-1}({h^{(\beta\gamma\delta)}}^{g_{(\alpha\beta)}})^{-1}h^{(\alpha\beta\gamma)}
\Delta^{(\alpha\gamma\delta)}_i
h^{(\alpha\beta\gamma)}{}^{-1}{h^{(\beta\gamma\delta)}}^{g_{(\alpha\beta)}} h^{(\alpha\beta\delta)}
+
\nn \\
&+
h^{(\alpha\beta\delta)}{}^{-1}({h^{(\beta\gamma\delta)}}^{g_{(\alpha\beta)}})^{-1}
\Delta^{(\alpha\beta\gamma)}_i
{h^{(\beta\gamma\delta)}}^{g_{(\alpha\beta)}} h^{(\alpha\beta\delta)}
+
h^{(\alpha\beta\delta)}{}^{-1}
\Delta^{(\beta\gamma\delta)}_i{}^{g_{(\alpha\beta)}}
h^{(\alpha\beta\delta)}
+
\Delta^{(\alpha\beta\delta)}_i
+
\nn \\
&-
h^{(\alpha\beta\delta)}{}^{-1}({h^{(\beta\gamma\delta)}}^{g_{(\alpha\beta)}})^{-1}
\tilde{a}_{(\alpha\beta)}
{h^{(\beta\gamma\delta)}}^{g_{(\alpha\beta)}} h^{(\alpha\beta\delta)}
+
h^{(\alpha\beta\delta)}{}^{-1}
\tilde{a}_{(\alpha\beta)}
h^{(\alpha\beta\delta)}
= 0,
\label{consistency-2}
\ea
where
\ba
{\Delta^{(\beta\gamma\delta)}_i}^{g_{(\alpha\beta)}}
:= g_{(\alpha\beta)}\Delta^{(\beta\gamma\delta)}_i g_{(\alpha\beta)}^{-1}.
\ea
In the Abelian case, this consistency condition can be proven to be a consequence 
of the consistency condition (\ref{consistency-1}),
which is independent of any information about $g^{(\alpha\beta)}$
or other fields besides $h^{(\alpha\beta\gamma)}$.
One can claim that,
given a set of $h^{(\alpha\beta\gamma)}$
satisfying the consistency condition (\ref{consistency-1}),
one can always find $g^{(\alpha\beta)}$ and $\ta^{(\alpha\beta)}$
to satisfy (\ref{new-consistency1}).
Furthermore, one can set $\bA^{(\alpha)} = 0$
and find $\tilde{B}^{(\alpha)}$ consistently on all patches.

However, in the non-Abelian case,
since the consistency condition (\ref{consistency-1})
has explicit dependence on $g^{(\alpha\beta)}$,
we cannot impose consistency conditions on
$h^{(\alpha\beta\gamma)}$
without giving $g^{(\alpha\beta)}$ first.
The information we need to construct a geometric structure which admits
a consistent solution of the potential 
$\tilde{B}^{(\alpha)}$
satisfying 
(\ref{Bg}) for $\Phi = \tilde{B}$ on all intersections
$U_{(\alpha\beta)}$ of two patches is
an assignment of 
$g^{(\alpha\beta)}$ and $\bA^{(\alpha)}$
such that
$h^{(\alpha\beta\gamma)}$ defined by (\ref{new-consistency1}) belongs to $H$
on all intersections $U_{(\alpha\beta\gamma)}$ of three patches.
Once $g^{(\alpha\beta)}$'s and $\bA^{(\alpha)}$'s are all given,
$\tilde{a}^{(\alpha\beta)}$ is uniquely fixed
by (\ref{Bg}) for $\Phi = \bA$, 
and the consistency condition (\ref{consistency-2}) is automatically satisfied.
In other words,
one has to first choose the geometric structure of the 1-form potential
before defining that of the 2-form potential.
It will be very interesting to find explicit examples
of topologically nontrivial non-Abelian gerbes.

\section{Conclusion}

In this note, we summarized the basic material about non-Abelian gerbes and
its application to M-theory.  In particular, we pointed out some of the difficulties
associated with the fake curvature condition
and a way out by modifying the gauge transformation for
some limited classes of crossed modules which
accept the existence of a map $s: G\rightarrow H$ with
certain constraints.

Obviously, there are many things to do in the future.
One of them is to understand the topologically nontrivial configurations
for such system as discussed in the last section.
Another important direction is to find more general classes of
gauge transformations which modify the non-Abelian gerbes
with better mathematical features.
In particular, at this moment, the fields which take values in $V$ 
interact only through the gauge potential corresponding to the gauge group $K$.  
We would like to know
if this will be a general feature. A hint toward such direction
has already been given in the analysis of \cite{r:MM5b}.

\section*{Acknowledgements}

YM would like to thank the organizers of the workshop,
``Mathematics and Applications of Branes in String and M-theory"
at Isaac Newton Institute at Cambridge, where he enjoyed
the communication with participants.
The authors would like
to thank N. Lambert, C. Saemann and M. Wolf
for their interest and discussions.
PMH is supported in part by
the National Science Council, Taiwan, R.O.C.
YM is partially supported by Grant-in-Aid
(KAKENHI \#20540253)
from MEXT Japan.

\appendix

\section*{Appendix}

In this appendix we introduce an alternative notation for 
the modified non-Abelian gerbes introduced in the paper.
Most equations are simplified in this new notation.

For the semi-direct product $G = K\ltimes V$,
we denote its Lie algebra generators as 
$\{T^i, T^a\}$,
where $T^i$'s and $T^a$'s are the generators of $\ttk$ and $\tth$, respectively.
The representation $\ur$ on $V$ defines a matrix
$\ur(T^i)^a{}_b$ for each generator $T^i$.
The Lie algebra $\ttg$ is then of the form
\ba
{}[T^i, T^j] &=& f^{ij}{}_k T^k, \\
{}[T^i, T^a] &=& \ur(T^i)^a{}_b T^b, \\
{}[T^a, T^b] &=& 0,
\ea
where $f^{ij}{}_k$ is the structure constant of $\ttk$.
In this notation, the maps $\ua, \ur$ are realized in terms of Lie brackets;
\ba
\ua({\bf x})(y) = [{\bf x}, y], \qquad \ur(x)(y) = [x, y]
\qquad
({\bf x}\in\ttg, \quad x\in\ttk \quad \mbox{and} \quad y\in\tth).
\ea
Similarly, the appearance of $\alpha, \rho$ can be avoided using
the following identities
\ba
\alpha(g)(y) = gyg^{-1}, \qquad
\rho(k)(y) = kyk^{-1}
\qquad
({\bf g}\in G, \quad k\in K \quad \mbox{and} \quad y\in\tth).
\ea
By identifying $V$ with a subalgebra of $\ttg$,
and the map $\ut$ is just an identity map and can be omitted.
The only two maps that have to be kept manifestly are $\us$ and $\Pi$.
The are the projections that maps an element in $\ttg$ 
to the part generated by $T^i$ or $T^a$,
respectively.
That is,
\ba
\us(T^i) = 0, \qquad \us(T^a) = T^a; \qquad
\Pi(T^i) = T^i, \qquad \Pi(T^a) = 0.
\ea

Let us rewrite some of the relations in the main text in the new notation.
In the new notation,
under an infinitesimal gauge transformation,
the one-form potential $A_i \in \ttg$ and
the two-form potential $B_{ij} \in \tth$ transform as
\ba
\delta \bA &=&
[\bD, \bL] + \tilde{a}, \\
\delta \tilde{B} &=&
\{\bD, \tilde{a}\}
- [\bL, \tilde{B}] + [\bF, \us(\bL)],
\ea
where
\ba
\bD = d + \bA
\ea
and the gauge transformation parameters are
$\bL \in \ttg$ and $\tilde{a}_i \in \tth$.

The gauge transformation algebra is given by
\ba
[\delta_{(1)}, \delta_{(2)}] = \delta_{(3)},
\ea
where the transformation parameters for $\delta_{(3)}$ are
\ba
\bL_{(3)} &=& [\bL_{(1)}, \bL_{(2)}], \\
\tilde{a}_{(3)} &=&
[\bL_{(1)}, \tilde{a}_{(2)}]
- [\bL_{(2)}, \tilde{a}_{(1)}].
\ea
In the new notation, the field strengths are
\ba
\bF &=&
\bD^2 - \tilde{B},
\\
\tilde{Z} &=&
[\bD, \tilde{B}] - [\bF, \us(\bA)]
\ea
and they transform covariantly as
\ba
\delta \bF &=&
[\bF, \Pi(\bL)],
\\
\delta \tilde{Z} &=&
[\tilde{Z}, \bL].
\ea

For finite gauge transformations
parametrized by $g \in G$ and $\tilde{a} \in \tth$,
the gauge potentials transform according to
\ba
\bD' &=&
g (\bD + \tilde{b}) g^{-1} = g \bD g^{-1} + \tilde{a}, 
\label{Dp}
\\
\tilde{B}' &=&
g \, \us\left( (\bD + \tilde{b})^2 \right) g^{-1}
- \us\left( g \bF g^{-1}\right).
\label{Bp}
\ea
where 
\ba
\tilde{b} := g^{-1}\tilde{a}g.
\ea
In the limit of infinitesimal gauge transformations,
$\tilde{b} = \tilde{a}$.

The composition of two consecutive finite gauge transformations by 
$(g_{(1)}, \tilde{a}_{(1)})$ and $(g_{(2)}, \tilde{a}_{(2)})$ 
is equivalent to a gauge transformation
by $(g_{(3)}, \tilde{a}_{(3)})$ with
\ba
g_{(3)} = g_{(1)} g_{(2)}, \qquad
\tilde{a}_{(3)} = \tilde{a}_{(1)} + g_{(1)}\tilde{a}_{(2)}g_{(1)}^{-1}.
\label{composition}
\ea
The gauge symmetry of the finite gauge transformations is
an equivalence relation
\ba
(g, \tilde{a}) \stackrel{h}{\approx} (g', \tilde{a}')
\label{ga}
\ea
with
\ba
g' &=& h g,
\label{ghg}
\\
\tilde{a}' &=& \tilde{a} + h^{-1}[\bD', h]
\label{gpap}
\ea
for arbitrary $h \in H$.
The finite gauge transformations of the field strengths are
\ba
\bF' &=& g^{-1} \bF g + \us(g^{-1}\Pi(F)g),
\\
\tilde{Z} &=& g^{-1}\tilde{Z}g.
\ea


\end{document}